\documentclass[12pt]{iopart}                  
\usepackage{iopams}                           
\newcommand{\text}[1]{\hbox{#1}}              
\newcommand{\bss}{\mathbf}                    
\newcommand{\D}{\rmd}                         
\newcommand{\gr}{{\rm\scriptstyle g}}
\newcommand{\g}{\hbox{\sl g}}
\newcommand{\Cdot}{\,}
\newcommand{\R}[1]{{\bi #1}}
\newcommand{\B}[1]{\protect{\bss #1}}
\newcommand{\VV}{\protect{\bss V}}
\newcommand{\UU}{\protect{\bss U}}
\begin{document}

\title{Turns and special relativity transformations}

\author{A A Ketsaris\footnote{{\it E-mail address\/}:
ketsaris@sai.msu.su}}
\address{19-1-83, ul. Krasniy Kazanetz, Moscow 111395, Russian Federation}

\date{\today}
\jl{1}

\begin{abstract}

We advance an universal approach to the construction of kinematics in
non-inertial and, in particular, rotating reference frames.  On its
basis, a 10-dimensional space including three projections of velocity
vector and three turn angles in geometric space as additional
coordinates to time and geometric coordinates is introduced.  In a
specific case, the turns in 10-space describe the uniform rotation of
reference frame as well as its accelerated motion.  Transformations for
coordinates and angular velocities are derived.  Definitions of dynamic
quantities contain a fundamental constant $\Omega$ with dimensionality
of angular velocity and the maximal acceleration in addition to the
velocity of light.  The special relativity relation between energy,
impulse, and mass gets changed for particles with moment of inertia.  A
wave equation obtained describes the rotary and accelerated motions of
light wave.  A relation between particle moment and the Planck constant
representing Bohr's postulate is added to de Broglie's relations.  A
generalized space-time which allows to consider kinematics for
derivations of arbitrary order is studied. Differential relations
between velocity of arbitrary order and turn angles in the generalized
space-time are obtained.

\end{abstract}

\section{Introduction}

It is well known that the non-relativistic view of global disk rotation
with constant angular velocity $\omega$ is in contradiction with the
relativistic rule of composition of velocities. It is believed that the
rotating reference frame can not be realized at distances larger than
$c/\omega$ as the rotation velocity should not exceed the velocity of
light. Despite it the global rotation model is widely used in the
description of relativistic effects in rotating frames such as the
Sagnac effect \cite{Sag}, the length contraction, the time dilation,
and others \cite{Sel97,Kla,Riz98,Tar99}. In doing so it is overlooked
that the global disk rotation is accelerated and rotary local motions
of disk parts related to each other.

Geometrical turns and Lorentz busts, i.e. turns in the
pseudoplane $(x,t)$, make full Lorentz group. The derivations of turn
parameters by time are angular velocity $\D\theta/\D t$ and
acceleration $\D v/\D t$.  It is a serious reason for considering both
angular velocity and acceleration in the unified way.

The progress has been possible in understanding relativistic laws for
accelerated motion due to the Caianiello's idea of the expansion of
space-time through components of four-velocity of particle
\cite{Cai80}.  The more important consequence of this model is the
existence of the maximal acceleration for any physical processes
\cite{Cai81,Sca}. In our recent work \cite{Ket} it was shown that a
similar increase of space-time dimension (through three components of
usual velocity) is required from an uniform approach to differentiation
of kinematic quantities. Namely, a derivation of one kinematic variable
by other should correspond to a turn angle in plane of these variables.
This is possible only if both the function-variable and the
argument-variable are involved in the construction of vector space.  So
in the special relativity theory, the velocity is in correspondence
with a turn angle, $\psi$, in the pseudoplane $(x,t)$:
\begin{equation}
     \frac{\D x}{\D t} = c\Cdot\tanh \psi \,.
\label{F0}
\end{equation}
By analogy with constant velocity motions, accelerated motions can be
considered through correspondence between the acceleration and the turn
angle in the pseudoplane $(v,t)$ like \eref{F0}.  Therefore the
velocity should be added as a new freedom degree of an expanded
space-time.

We shall now present the aforesaid in more rigorous form.
Let us write an interval square used in the special relativity
as
\[
     (\D s)^2 = - (\D x^1)^2 - (\D x^2)^2 - (\D x^3)^2 + c^2\Cdot (\D t)^2 \,.
\]
Turns in the pseudoplanes $(\D x^a,\D t)$, where
$a=1,2,3$, preserve the interval square.  The parameters
of these turns are the velocity coordinates $v^a=\D x^a/\D t$.
In our previous work \cite{Ket} the following logical scheme was proposed
\begin{enumerate}
\item to supplement the differentials $\D x^a$ and $\D t$ by the
differentials of turn parameters $\D v^a$;

\item to construct a vector space on these differentials;

\item to introduce the interval square in the specified vector space;

\item to consider turns in the pseudoplanes $(\D v^a,\D t)$.
\end{enumerate}
As a result, it is possible to generalize the special relativity to
motions with variable velocity.

The geometric rotations in the planes $(\D x^1,\D x^2)$, $(\D x^3,\D x^1)$,
$(\D x^2,\D x^3)$ preserve also the interval square.  The parameters of
these turns are angles $\theta_1^2$, $\theta_3^1$, $\theta_2^3$,
respectively.  According to the above approach, a kinematic description
of geometric rotations with angular velocity must result from a relation
between turns angles in the planes $(\D \theta_1^2, \D t)$, $(\D
\theta_3^1, \D t)$, $(\D \theta_2^3, \D t)$ and angular velocity
components. Thus the geometrical angles $\theta_1^2$, $\theta_3^1$,
$\theta_2^3$ should be included in a list of freedom degrees of space.

The present paper is aimed at applying these ideas to geometric
rotations and at constructing the kinematics for derivations of arbitrary
order.

The rotation kinematics, including rules of composition of angular
velocities and transformations of kinematic variables, is considered
in Section~2.  A generalization of dynamic variables (energy, impulse,
force, moment) is produced in Section~3 where a wave equation is also
modified to describe rotary motion of light wave.  Section~4 contains
a special relativity generalization to the kinematics of arbitrary order.  The
conclusions are presented in Section~5.

\section{Transformations for rotating reference frames}

We supplement the differentials $\D x^a$, $\D t$ by the angle
differentials $\D \theta_1^2$, $\D \theta_3^1$, $\D \theta_2^3$ to
construct a vector space on them.  Let us introduce an interval square
\[
     \fl
     (\D s)^2 = - (\D x^1)^2 - (\D x^2)^2 - (\D x^3)^2 + c^2\Cdot (\D t)^2
     {}+
     R^2\Cdot (\D \theta_1^2)^2 + R^2\Cdot (\D \theta_3^1)^2 +
     R^2\Cdot (\D \theta_2^3)^2
\]
in this vector space.
Here the constant $R$ adjusts the dimensionality of angles $\D \theta_a^b$
to the dimensionality of interval and has the dimensionality of length.
We invoke the fundamental time $T$ and the fundamental length $L=c\Cdot T$
introduced in \cite{Ket}.
Let the fundamental length be related to $R$ by
\[
      L = 2\Cdot \pi \Cdot R \Cdot N \,,
\]
where $N$ is the dimensionless parameter.
If we divide the interval square by $L^2$, we obtain
\begin{eqnarray*}
     (\D \sigma )^2 \equiv \frac{(\D s)^2}{L^2} =
     &-& \frac{1}{L^2} \Cdot
     \left[ (\D x^1)^2 + (\D x^2)^2 + (\D x^3)^2\right]
     + \frac{(\D t)^2}{T^2} \\
     &+& \frac{1}{4\Cdot \pi^2\Cdot N^2} \Cdot \left[
       (\D \theta_1^2)^2 + (\D \theta_3^1)^2 + (\D \theta_2^3)^2 \right] \,.
\end{eqnarray*}
Let us change variables
\[
     x^a   = \frac{x^a}{L}     \,,\qquad
     x^4   = \frac{t}{T}\,,\qquad
     x_a^b = \frac{\theta_a^b}{2\Cdot \pi\Cdot N}\,,
\]
where the indexes $a$ and $b$ numbering geometric coordinates take
values $1,2,3$.
Then the interval square can be rewritten in the dimensionless form
\[
     \fl
     (\D \sigma )^2 = - (\D x^1)^2 - (\D x^2)^2 - (\D x^3)^2
     + (\D x^4)^2 + (\D x_1^2)^2  + (\D x_3^1)^2  + (\D x_2^3)^2 \,.
\]

\subsection{$\bTheta\Cdot\bOmega$-motion of reference frame}

We consider a body $B$ in uniform rotation with respect to a frame $K$
in the plane $(\D x^1, \D x^2)$ about the angle $\theta$ with the
angular velocity $\omega$.  Let a frame $K'$ rotate uniformly with
respect to $K$ in the same plane about the angle $\btheta$ with the
angular velocity $\bomega$.  The motion of body $B$ will be
characterized by the angular velocity $\omega'$ and the angle $\theta'$
with respect to $K'$.  In this case the dimensionless interval square
is simplified
\[
     (\D \sigma )^2 = - (\D x^1)^2 - (\D x^2)^2 + (\D x^4)^2 + (\D x_1^2)^2 \,.
\]

A turn in the space of differentials
\[
     ||\D x|| = \UU\Cdot ||\D x'||
\]
preserves the interval square.
The turn of frame in the plane $(\D x^1, \D x^2)$ about the angle $\btheta$
is given by the matrix
\[
\bTheta =
\begin{array}{|c|c|c|c|}
\hline
\cos\btheta & -\sin\btheta  &    &  \\
\hline
\sin\btheta & \cos\btheta   &    &  \\
\hline
         &                & 1  &  \\
\hline
         &                &    & 1\\
\hline
\end{array} \,.
\]
The turn in the plane $(\D x^4,\D x_1^2)$ defined by the matrix
\[
\bOmega =
\begin{array}{|c|c|c|c|}
\hline
 1 &  &              &              \\
\hline
   &1 &              &              \\
\hline
   &  & \cos\bxi  & -\sin\bxi  \\
\hline
   &  & \sin\bxi & \cos\bxi \\
\hline
\end{array}
\]
corresponds uniform rotation of frame. The turns
$\bTheta$ and $\bOmega$ are commutating to one another.
The sequential realization of these turns gives 
the linear transformation
\begin{equation}
\eqalign{
     \D x^1   &= \cos\btheta \Cdot (\D x^1)'
     {}- \sin\btheta\Cdot (\D x^2)'\,,\\
     \D x^2   &= \sin\btheta \Cdot (\D x^1)'
     {}+ \cos\btheta\Cdot (\D x^2)'\,,\\
     \D x^4   &= \cos\bxi \Cdot (\D x^4)'
     {}- \sin\bxi \Cdot (\D x_1^2)' \,,\\
     \D x_1^2 &= \sin\bxi \Cdot (\D x^4)'
     {}+ \cos\bxi \Cdot (\D x_1^2)' \,.
}
\label{F1}
\end{equation}

In order to find a relation between the angle $\bxi$ of
turn matrix $\bOmega$
and the angular velocity of body $B$, we consider the variations of
coordinate differentials as functions of the turn angle variations:
\[
     \delta ||\D x|| = \delta \UU\Cdot ||\D x'|| \,.
\]
Taking into account that
\[
     ||\D x'|| = \UU^{-1}\Cdot ||\D x|| \,,
\]
we get
\[
     \delta ||\D x|| = (\delta \UU\Cdot \UU^{-1})\Cdot ||\D x|| \,.
\]
In our case $\UU =\bTheta\Cdot\bOmega = \bOmega\Cdot\bTheta$ and
\[
     \delta \UU\Cdot \UU^{-1} =
     \delta \bTheta\Cdot \bTheta^{-1} +
     \delta \bOmega\Cdot \bOmega^{-1}\,.
\]
Using this matrix we obtain
\begin{equation}
\eqalign{
     \delta \D x^1   = - \delta \btheta \Cdot \D x^2  \,,\qquad&
     \delta \D x^4   = - \delta \bxi \Cdot \D x_1^2   \,,\\
     \delta \D x^2   =   \delta \btheta \Cdot \D x^1  \,,\qquad&
     \delta \D x_1^2 =   \delta \bxi\Cdot \D x^4      \,.
}
\label{F2}
\end{equation}
Consider the differential
\[
     \delta \frac{\D x_1^2}{\D x^4} = \frac{\delta \D x_1^2}{\D x^4} -
     \D x_1^2 \Cdot \frac{\delta \D x^4}{(\D x^4)^2}\,.
\]
From (\ref{F2}) follows
\begin{equation}
     \delta x_{14}^2 = \delta \bxi\Cdot \left[ 1 + (x_{14}^2)^2 \right]\,,
\label{F3}
\end{equation}
where the notation $x_{14}^2=\D x_1^2/\D x^4$ was introduced.

The relations \eref{F1} and \eref{F3} determine
the coordinate transformations and the rule of composition of
angular velocities in the case of uniform rotation of reference frame.

\subsubsection{Rule of composition of angular velocities}

As the rotations $\bTheta$ and $\bOmega$ are commutating, the
rule of composition of angular velocities does not depend on turn angle of
frame.
From \eref{F3} one can obtain
\[
     x_{14}^2 = \tan (\bxi + \xi')\,,
\]
where $\xi'$ is integration constant. Let us take the relations
$\R{x}_{14}^2=0$ for $\bxi=0$, $(x_{14}^2)'=0$ for $\xi'=0$ as
initial conditions. Then
\begin{equation}
     \R{x}_{14}^2 = \tan \bxi
\label{F4}
\end{equation}
and $(x_{14}^2)'=\tan\xi'$. Thus we find
the rule of composition of angular velocities
\[
     x_{14}^2 =
     \frac{\R{x}_{14}^2 + (x_{14}^2)'}{1-\R{x}_{14}^2\Cdot (x_{14}^2)'} \,.
\]
If we change for dimensional values in correspondence with
\[
     x_{14}^2 = \frac{\D x_1^2}{\D x^4} =
     \frac{T}{2\Cdot \pi\Cdot N}\Cdot \frac{\D\theta_1^2}{\D t} =
     \frac{\omega}{\Omega}                  \,,
\]
where the notation
\[
     \Omega=\frac{2\Cdot \pi\Cdot N}{T}=\frac{c}{R}
\]
was introduced, we obtain
\[
     \omega = \frac{\bomega + \omega'}{1-\frac{\bomega\Cdot \omega'}
     {\Omega^2}} \,.
\]
As we see, the constant $\Omega$ is a fundamental angular velocity
like the light velocity in the rule of composition of velocities. The
relation for composition of angular velocities has a singularity when
\[
     \bomega\Cdot \omega' = \Omega^2 \,.
\]
It is possible that the presence of this singularity
has a profound physical meaning.

\subsubsection{Transformations of differentials of coordinates}

From \eref{F4} follows
\[
     \cos\bxi =
     \frac{1}{\sqrt{1+\frac{\bomega^2}{\Omega^2}}} \,, \qquad
     \sin\bxi =
     \frac{\bomega}{\Omega\Cdot \sqrt{1+\frac{\bomega^2}{\Omega^2}}}\,.
\]
If we substitute the above expressions in \eref{F1}
and change for dimensional values,
we obtain the transformations of coordinate differentials:
\begin{equation}
\eqalign{
     \D x^1   &= \cos\btheta \Cdot (\D x^1)'
     {}- \sin\btheta\Cdot (\D x^2)'\,,\\
     \D x^2   &= \sin\btheta \Cdot (\D x^1)'
     {}+ \cos\btheta\Cdot (\D x^2)'\,,\\
     \D t     &=
     \frac{1}{\sqrt{1+\frac{\bomega^2}{\Omega^2}}} \Cdot (\D t)' -
     \frac{\bomega}{\Omega^2\Cdot
        \sqrt{1+\frac{\bomega^2}{\Omega^2}}} \Cdot (\D \theta)' \,,\\
     \D \theta &=
     \frac{\bomega}{\sqrt{1+\frac{\bomega^2}{\Omega^2}}} \Cdot (\D t)'
     {}+ \frac{1}{\sqrt{1+\frac{\bomega^2}{\Omega^2}}}
     \Cdot (\D \theta)' \,.
}
\label{F5}
\end{equation}

\section{Relativistic mechanics}

The necessity of the correction of relativistic mechanics stems from
the fact that the velocities $v^a$ and the angles $\theta_b^a$ of
geometrical turns are introduced as additional coordinates being
independent of geometrical coordinates and time. The angles
$\theta_b^a$ can be identified with particle interior freedom degrees
responsible for a particle {\it spin}.  Let us construct the relativistic
mechanics generalized to accelerated motions and uniform rotations by
analogy to the relativistic mechanics to be invariant with respect to
uniform velocity motions.

\subsection{10-velocity}

The differentials included in the interval square
\begin{eqnarray*}
     \fl
     (\D s)^2 &=& c^2\Cdot (\D t)^2 - (\D x^1)^2 - (\D x^2)^2 - (\D x^3)^2 \\
     \fl
     &&{}-
     T^2\Cdot (\D v^1)^2 - T^2\Cdot (\D v^2)^2 - T^2\Cdot (\D v^3)^2
     {}+
     R^2\Cdot (\D \theta^3_2)^2 + R^2\Cdot (\D \theta^1_3)^2 +
     R^2\Cdot (\D \theta^2_1)^2 \,,
\end{eqnarray*}
can be considered as the vector coordinates in 10-dimensional space.
We have for the contravariant coordinates
\[
     \fl
     \D x^\mu = \{ c\Cdot \D t ,\;  \D x^1 ,\;  \D x^2 ,\;  \D x^3 ,\;
     T\Cdot \D v^1 ,\; T\Cdot \D v^2 ,\; T\Cdot \D v^3 ,\;
     R\Cdot \D \theta^3_2 ,\; R\Cdot \D \theta^1_3 ,\;
     R\Cdot \D \theta^2_1 \}
\]
and for the covariant coordinates
\[
     \fl
     \D x_\mu = \{c\Cdot \D t ,\; - \D x^1 ,\; - \D x^2 ,\;  - \D x^3 ,\;
     - T\Cdot \D v^1 ,\;  - T\Cdot \D v^2 ,\;  - T\Cdot \D v^3 ,\;
     R\Cdot \D \theta^3_2 ,\;  R\Cdot \D \theta^1_3 ,\;
     R\Cdot \D \theta^2_1\}\,.
\]
The interval square is rewritten in the new coordinates as
\[
   (\D s)^2 = \D x^\mu \Cdot \D x_\mu \,,
   \qquad\qquad(\mu=1,\ldots,10) \,.
\]
Let us introduce a {\it generalized Lorentz factor}
\[
    \gamma =  \left(
        1-\frac{v^2}{c^2}-\frac{a^2}{A^2}+\frac{\omega^2}{\Omega^2}
    \right)^{-1/2} \,.
\]
We express the interval as
\[
     \D s = \frac{c\Cdot \D t}{\gamma}
\]
and define a {\it 10-velocity} as
\[
     u^\mu \equiv \frac{\D x^\mu}{\D s} =
     \frac{\partial s}{\partial x_\mu} =
     \left\{ \gamma \,,\; \gamma\Cdot \frac{v^b}{c} \,,\;
             \gamma\Cdot \frac{a^b}{A} \,,\;
             \gamma\Cdot \frac{\omega^b_a}{\Omega}
     \right\}
\]
in contravariant coordinates, and
\[
     u_\mu \equiv \frac{\D x_\mu}{\D s} =
     \frac{\partial s}{\partial x^\mu} =
     \left\{ \gamma \,,\; -\gamma\Cdot frac{v_b}{c} \,,\;
             -\gamma\Cdot frac{a_b}{A} \,,\;
             \gamma\Cdot \frac{\omega^a_b}{\Omega}
     \right\}
\]
in covariant coordinates.
For this $v^b=v_b$, $a^b=a_b$, $\omega^b_a=\omega^a_b$,
$v^b\Cdot v_b=v^2$, $a^b\Cdot a_b=a^2$, $\omega^b_a\Cdot
\omega^a_b=\omega^2$.
It is obvious that
\begin{equation}
     u^\mu\Cdot u_\mu = 1 \,.
\label{F6}
\end{equation}

\subsection{Operation for free particle}

Let us define an {\it operation} for free particle with mass and moment of
inertia as
\[
     S = - \frac{E_0}{c}\Cdot \int\limits_{s_1}^{s_2} \D s\,,
\]
where the integration is over a line in 10-space, $s_1$, $s_2$ are points of
the specified line,
\[
     E_0 = m\Cdot c^2 = J\Cdot \Omega^2\,
\]
is the rest energy of particle, and $J=m\Cdot R^2$. The quantity $J$ will be
called {\it proper moment of inertia} of particle.
The operation can be expressed as an integral over time
\[
     S = - E_0\Cdot \int\limits_{t_1}^{t_2}
     \sqrt{1-\frac{v^2}{c^2}-\frac{a^2}{A^2}+\frac{\omega^2}{\Omega^2}}
     \Cdot \D t\,.
\]

\subsection{Energy, impulse, force, moment}

We define a {\it 10-impulse} as
\[
     p^\mu \equiv - \frac{\partial S}{\partial x_\mu} =
     \frac{E_0}{c} \Cdot u^\mu
     \qquad \text{and} \qquad
     p_\mu \equiv - \frac{\partial S}{\partial x^\mu} =
     \frac{E_0}{c} \Cdot u_\mu
\]
in contravariant and covariant coordinates, respectively.
Let us introduce a {\it relativistic energy}
\[
    E = \gamma\Cdot m\Cdot c^2 \,,
\]
a {\it relativistic impulse}
\[
    p=\gamma\Cdot m\Cdot v \,,
\]
a {\it relativistic kinetic force}
\[
    f = \gamma\Cdot m\Cdot a \,,
\]
and a {\it relativistic moment}
\[
    l = \gamma\Cdot J\Cdot \omega \,.
\]
According to our concept, the relativistic moment being independent of
particle motion in geometrical space can be identified with the spin of
particle.

Through quantities introduced the component of 10-impulse can be written as
\[
     \fl
     p^\mu = \left\{ \frac{E}{c} \,,\; p^b \,,\; T\Cdot f^b \,,\;
     \frac{l^b_a}{R}\right\}
     \qquad \text{and} \qquad
     p_\mu = \left\{ \frac{E}{c} \,,\; - p_b \,,\; - T\Cdot f_b \,,\;
      \frac{l^a_b}{R}\right\} \,.
\]
From \eref{F6} follows
\[
     p^\mu\Cdot p_\mu = \frac{(E_0)^2}{c^2}\,.
\]
This can be written as a relation between energy, impulse, force, moment and
rest energy in the relativistic mechanics generalized to accelerated motions
and uniform rotations:
\[
     \frac{E^2}{c^2} - p^2 - f^2\Cdot T^2 +
     \frac{l^2}{R^2}=\frac{(E_0)^2}{c^2}\,.
\]
For particles with zero rest energy, we have
\begin{equation}
     \frac{E^2}{c^2} - p^2 - f^2\Cdot T^2 + \frac{l^2}{R^2}=0\,.
\label{F7}
\end{equation}

The transformation of components of 10-impulse can be described by
the formalism similar to that for transformation of
coordinate differentials. For example, in the case of
$\bTheta\Cdot\bOmega$-motion of frame, the transformations \eref{F5} imply
\begin{eqnarray*}
     p^1   &=& \cos\btheta \Cdot (p^1)'
     {}- \sin\btheta\Cdot (p^2)'\,,\\
     p^2   &=& \sin\btheta \Cdot (p^1)'
     {}+ \cos\btheta\Cdot (p^2)'\,,\\
     E     &=&
     \frac{1}{\sqrt{1+\frac{\bomega^2}{\Omega^2}}} \Cdot E' -
     \frac{\bomega}{
        \sqrt{1+\frac{\bomega^2}{\Omega^2}}} \Cdot l' \,,\\
     l     &=&
     \frac{\bomega}{\Omega^2\Cdot\sqrt{1+\frac{\bomega^2}{\Omega^2}}}
     \Cdot E'
     {}+ \frac{1}{\sqrt{1+\frac{\bomega^2}{\Omega^2}}} \Cdot l' \,.
\end{eqnarray*}

\subsection{Wave equation}

A 10-dimensional derivative operator is given by
\[
     \frac{\partial }{\partial x_\mu} = \left\{
     \frac{1}{c}\frac{\partial }{\partial t} \,,\;
     -\frac{\partial }{\partial x_b} \,,\;
     -\frac{1}{T}\frac{\partial }{\partial v_b} \,,\;
     \frac{1}{R}\frac{\partial }{\partial \theta^a_b}\right\}
\]
for contravariant components and by
\[
     \frac{\partial }{\partial x^\mu} = \left\{
     \frac{1}{c}\frac{\partial }{\partial t} \,,\;
     \frac{\partial }{\partial x^b} \,,\;
     \frac{1}{T}\frac{\partial }{\partial v^b} \,,\;
     \frac{1}{R}\frac{\partial }{\partial \theta^b_a}\right\}
\]
for covariant components.
Using the derivative operator one can write a wave equation for light
\begin{equation}
     \frac{\partial^2 \psi}{\partial x^\mu \Cdot \partial x_\mu}\equiv
     \frac{1}{c^2}\frac{\partial^2 \psi}{\partial t^2}
     -\frac{\partial^2 \psi}{\partial x^2} -
     \frac{1}{T^2}\frac{\partial^2 \psi}{\partial v^2}+
     \frac{1}{R^2}\frac{\partial^2 \psi}{\partial \theta^2}=0 \,.
\label{F8}
\end{equation}
This wave equation is different from the conventional one by two later
addends allowing to describe accelerated light motion as well as uniform
rotary light motion.

Let $\psi(t,x,v,\theta)$ be an arbitrary function describing
the wave field. We shall try for the solution of wave equation \eref{F8}
in the form
\begin{equation}
     \psi = \psi_0 \Cdot\exp\left[i\left(
     \kappa_b\Cdot x^b + \xi_b\Cdot v^b - \nu\Cdot t -
     n_b^a\Cdot\theta^b_a\right)\right]\,,
\label{F9}
\end{equation}
where $\kappa_b$ are the coordinates of wave vector, $\xi_b$ are the
coordinates of wave velocity vector \cite{Ket},
$\nu$ is the wave {\it circular} frequency.
The quantities $n_b^a$ will be called coordinates of {\it vector of phase
multiplicity}.  The length of this vector $n=\sqrt{n_b^a\Cdot n_a^b}$ will be
called {\it phase multiplicity}.
After substitution \eref{F9} in the wave equation, we get
\begin{equation}
     \frac{\nu^2}{c^2} - \kappa^2 -
     \frac{\xi^2}{T^2} + \frac{n^2}{R^2}=0\,.
\label{F10}
\end{equation}
If we multiply this equation by the Planck constant square and
compare the result to \eref{F7} within the framework
of corpuscular-wave duality, we obtain a set of relations for
particles with zero rest energy:
\[
     E = \hbar\Cdot\nu \,, \quad
     p = \hbar\Cdot\kappa    \,, \quad
     f =\frac{\hbar}{T^2} \Cdot \xi\,,\quad
     l = \hbar\Cdot n \,.
\]
The first two are de Broglie's relations; the third one is the relation
between the wave vector of velocity $\xi$ and the relativistic kinetic
force; the last one represents the {\it Bohr postulate}
generalized to particles with zero rest energy.

By analogy with the traditional definition of wave velocity
\[
     v^b \equiv \frac{\partial \nu}{\partial \kappa_b}
     = \frac{\kappa^b}{\nu}\Cdot c^2
\]
and the wave acceleration defined in \cite{Ket} by
\[
     a^b \equiv \frac{\partial \nu}{\partial \xi_b}
     = \frac{\xi^b}{\nu}\Cdot A^2 \,,
\]
we define a {\it wave angular velocity}
\[
     \omega_a^b \equiv \frac{\partial \nu}{\partial n^a_b}
     = \frac{n_a^b}{\nu}\Cdot \Omega^2\,.
\]
From (\ref{F10}) we obtain the relation
\[
     c^2 - v^2 - a^2\Cdot T^2 + \omega^2\Cdot R^2 = 0\,,
\]
which will be called {\it equation of light motion}.

The wave equation generalizing the Klein-Gordon equation for massive
particles to accelerated motion can be written as
\[
     \frac{\partial^2\psi}{\partial x^\mu \Cdot \partial x_\mu}
     + \frac{m^2 \Cdot c^2}{\hbar^2}\psi = 0\,.
\]
If the solution of this equation looks like the function \eref{F9}, then
\begin{equation}
     \frac{\nu^2}{c^2} - \kappa^2 -\frac{\xi^2}{T^2} + \frac{n^2}{R^2}=
     \frac{m^2 \Cdot c^2}{\hbar^2}\,.
\label{F12}
\end{equation}
Let us assume
\[
     \nu = \frac{m\Cdot c^2}{\hbar} + \nu'\,,
\]
where $\nu'\ll \frac{m\Cdot c^2}{\hbar}$.
For frequency $\nu'$ the equation \eref{F12} is reduced to
\begin{equation}
    \frac{2\Cdot m\Cdot \nu'}{\hbar} - \kappa^2 -\frac{\xi^2}{T^2}
    + \frac{n^2}{R^2}= 0\,.
\label{F13}
\end{equation}
By analogy to the traditional definition of wave packet
{\it group velocity}
\[
     v^b_{\gr} \equiv \frac{\partial \nu'}{\partial \kappa_b} =
     \frac{\hbar}{m}\Cdot \kappa^b
\]
and the wave packet group acceleration defined in \cite{Ket} by
\[
     a^b_{\gr} \equiv \frac{\partial \nu'}{\partial \xi_b} =
     \frac{\hbar}{m}\Cdot \frac{\xi^b}{T^2} \,,
\]
we define a wave packet {\it group angular velocity} as
\[
     (\omega^b_a)_{\gr} \equiv \frac{\partial \nu'}{\partial n^a_b} =
     \frac{\hbar}{m}\Cdot \frac{n^b_a}{R^2}\,.
\]
From (\ref{F13}) we obtain the expression
\[
     \hbar\Cdot\nu'  = \frac{m\Cdot v_{\gr}^2}{2} +
     \frac{m\Cdot T^2\Cdot a_{\gr}^2}{2} -
     \frac{m\Cdot R^2\Cdot \omega_{\gr}^2}{2}\,,
\]
which is a generalization of the known de Broglie's relation.

\section{Special relativity generalization to kinematics of arbitrary order}

\subsection{Generalized space-time}

In the previous Sections was shown that if turn angles in the planes $(\D
x^{i_1},\D x^{i_2})$ are considered as additional coordinates $x^{i_1i_2}$ to
space-time coordinates $x^{i_1}$, the relativistic
kinematic for accelerated and rotary motions can be produced.  In the space
with such coordinates, the interval is represented by differentials
$\D x^{i_1}$, $\D x^{i_1i_2}$, and the dimensionless interval square has the
form
\[
     (\D \sigma )^2 = \g_{i_1k_1}\Cdot \D x^{i_1}\Cdot \D x^{k_1} +
     \g_{i_1i_2,k_1k_2}\Cdot \D x^{i_1i_2}\Cdot \D x^{k_1k_2} \,.
\]
The complexity of the algebraic structure of space is determined by
the maximal tensor degree of additional coordinates.
It will be called {\it kinematics order}.
To construct the kinematics of higher order, the above approach should be
applied by a recurrence way.

In the general case, the kinematics of order $n$ is considered in the space
with coordinates $x^{i_1i_2i_3i_4...i_k}$, where $k=1\ldots n$.
These coordinates are turn angles in the planes
$(\D x^{i_1i_2\ldots i_m},\D x^{i_mi_{m+1}...i_k})$, where $m=k/2,\ldots,k-1$.
The space itself will be called {\it generalized space-time} and
will be denoted by $\Bbb{X}_N$, where $N$ is its dimension.
In the generalized space-time, the interval square is written as
\begin{eqnarray*}
     (\D \sigma )^2 &=& \g_{i_1k_1}\Cdot \D x^{i_1}\Cdot \D x^{k_1} +
     \g_{i_1i_2,k_1k_2}\Cdot \D x^{i_1i_2}\Cdot \D x^{k_1k_2} \\
     &&{}+ \ldots +
     \g_{i_1i_2...i_n,k_1k_2...k_n}\Cdot
     \D x^{i_1i_2...i_n}\Cdot \D x^{k_1k_2...k_n} \,.
\end{eqnarray*}
A {\it collective index} $I$ taking values $i_1, i_1i_2,\ldots,i_1i_2...i_n$
can be defined as convenience. The interval square
is rewritten through the collective index in the easy-to-use form:
\begin{equation}
     (\D \sigma )^2 = \g_{IK}\Cdot \D x^I\Cdot \D x^K \,.
\label{F14}
\end{equation}
The interval square of this kind is a particular case of
generalizations made in our book \cite{Book}.

\subsection{Structure equations of the generalized space-time}

Let us consider a turn in the generalized space-time $\Bbb{X}_N$,
i.e. a linear transformation
\[
     ||\D x|| = \UU\Cdot ||\D x'||
\]
preserving the interval square \eref{F14}. The turn matrix $\UU$
depends on the turn angles $x^\alpha$ which are the coordinates of
$\Bbb{X}_{N+1}$.  A set of turns is group. In other words, for this set
a composition rule
\begin{equation}
     \UU = \UU_2\Cdot \UU_1
\label{F15}
\end{equation}
can be introduced, an unity element can be defined,
and an inverse element exists for any element from $\UU$.

Consider the variations of coordinate differentials as
functions of the turn angle variations:
\begin{equation}
     \delta ||\D x|| = \delta \UU\Cdot ||\D x'||
     = \Delta \UU\Cdot ||\D x|| \,,
\label{F16}
\end{equation}
where the notation $\Delta \UU = \delta \UU\Cdot \UU^{-1}$ was introduced.
These relations will be called {\it structure equations} of the generalized
space-time (for more details, see \cite{Book}). In the expanded form they are
written as
\[
     \delta \D x^I = {\Delta \UU(x^\alpha)^I}_K\Cdot \D x^K \,.
\]
In a special case the structure equations reduce to the equations
\eref{F2} and the equations (7), (12) from \cite{Ket}.

In the neighbourhood of the turn group unit, the structure equations
take the form
\[
     \delta \D x^I = {\B{C}^I}_{K\alpha}\Cdot\delta x^\alpha\Cdot \D x^K \,,
\]
where the notation
\[
     {\B{C}^I}_{K\alpha} =
     \left. \partial_\alpha {\UU^I}_K \right|_{x^\alpha=0}
\]
is used.
The values ${\B{C}^I}_{K\alpha}$ are called the {\it structure constants}.
In particular, if the set of turn angles is the same as the coordinate set
of $\Bbb{X}_N$, the structure equations is written as
\[
     \delta \D x^I = {\B{C}^I}_{KL}\Cdot\delta x^L\Cdot \D x^K \,.
\]

From the composition rule \eref{F15} it follows that
\[
     \delta \UU = \delta \UU_2\Cdot \UU_1 +
     \UU_2\Cdot \delta \UU_1\,.
\]
If we multiply this expression on the right by
$\UU^{-1} = (\UU_1)^{-1}\Cdot (\UU_2)^{-1}$ and 
substitute the result in \eref{F16},
we obtain the structure equations
for the turn angles involved in the composition rule:
\[
     \delta ||\D x|| = \left[ \Delta \UU_2 +
     \UU_2\Cdot \Delta \UU_1\Cdot (\UU_2)^{-1}
     \right] \Cdot ||\D x|| \,.
\]

\subsection{Relation between velocity of arbitrary order and generalized
space-time coordinates}

An arbitrary motion in the generalized space-time makes itself evident
in the fact that one coordinate set depends on other coordinate set.
In the simplest case, for example, the space coordinates are functions of
time: $x^a=x^a(x^4)$.

Consider a set of arguments, $x^{\alpha_1}$,
and  a set of functions, $x^{\alpha_2}$, in the
generalized space-time $\Bbb{X}_N$. In other words, it is
proposed that functional relationships $x^{\alpha_2}(x^{\alpha_1})$ exist.
The differentials of these functions can be written as
\begin{equation}
     \D x^{\alpha_2} = {\VV^{\alpha_2}}_{\beta_1}\Cdot \D x^{\beta_1}\,.
\label{F17}
\end{equation}
The quantity ${\VV^{\alpha_2}}_{\beta_1} \equiv
\partial_{\beta_1}x^{\alpha_2}$
will be called {\it velocity of arbitrary order}.
We assume that the remaining coordinates of $\Bbb{X}_N$
are invariant. Then the interval is represented by the differentials
$\D x^{\alpha_2}$, $\D x^{\alpha_1}$,
and the interval square has the form
\[
     (\D \sigma )^2 = \g_{\alpha\beta}\Cdot\D x^{\alpha}\Cdot\D x^{\beta}  =
     \g_{\alpha_2\beta_2}\Cdot\D x^{\alpha_2}\Cdot\D x^{\beta_2} +
     \g_{\alpha_1\beta_1}\Cdot\D x^{\alpha_1}\Cdot\D x^{\beta_1}\,.
\]

Introduce a turn in $\Bbb{X}_N$
\[
     \D x^{\alpha} = {\UU^{\alpha}}_\beta\Cdot (\D x^\beta)' \,.
\]
We shall find a relation between velocity of arbitrary order and
generalized space-time coordinates which are turn angles in the planes
$(\D x^{\alpha},\D x^\beta)$. For this purpose we consider the variations of
coordinate differentials:
\[
     \delta \D x^{\alpha} = {\Delta \UU^{\alpha}}_\beta\Cdot \D x^\beta \,,
\]
where
${\Delta \UU^{\alpha}}_\beta = {(\delta \UU\Cdot \UU^{-1})^{\alpha}}_\beta$.
This can be rewritten in the expanded form
\begin{equation}
\eqalign{
     \delta \D x^{\alpha_2} &=
     {\Delta \UU^{\alpha_2}}_{\beta_2}\Cdot \D x^{\beta_2}
     + {\Delta \UU^{\alpha_2}}_{\beta_1}\Cdot \D x^{\beta_1} \,,\\
     \delta \D x^{\alpha_1} &=
     {\Delta \UU^{\alpha_1}}_{\beta_2}\Cdot \D x^{\beta_2}
     + {\Delta \UU^{\alpha_1}}_{\beta_1}\Cdot \D x^{\beta_1} \,.
}
\label{F18}
\end{equation}
If we differentiate the equation \eref{F17} by the angles of turn matrix,
we obtain
\[
    \delta \D x^{\alpha_2} =
    \delta {\VV^{\alpha_2}}_{\beta_1}\Cdot \D x^{\beta_1}
    + {\VV^{\alpha_2}}_{\beta_1}\Cdot \delta \D x^{\beta_1}\,.
\]
From this expression, \eref{F17}, and \eref{F18} follows
\[
    \fl
    \delta {\VV^{\alpha_2}}_{\beta_1} = {\Delta \UU^{\alpha_2}}_{\beta_1} +
    {\Delta \UU^{\alpha_2}}_{\beta_2}\Cdot {\VV^{\beta_2}}_{\beta_1} -
    {\VV^{\alpha_2}}_{\gamma_1}\Cdot {\Delta \UU^{\gamma_1}}_{\beta_1}
    {}-
    {\VV^{\alpha_2}}_{\gamma_1}\Cdot {\Delta \UU^{\gamma_1}}_{\beta_2}
    \Cdot {\VV^{\beta_2}}_{\beta_1} \,.
\]
This differential equation establishes the desired relation between
velocity of arbitrary order and generalized space-time coordinates.

In the case when turns are considered only in planes
$(\D x^{\alpha_2},\D x^{\beta_1})$,
\[
     {\Delta \UU^{\alpha_2}}_{\beta_2} = 0 \,, \quad
     {\Delta \UU^{\alpha_1}}_{\beta_1} = 0 \,,
\]
and the differential equation obtained is simplified:
\begin{equation}
    \delta {\VV^{\alpha_2}}_{\beta_1} = {\Delta \UU^{\alpha_2}}_{\beta_1} -
    {\VV^{\alpha_2}}_{\gamma_1}\Cdot {\Delta \UU^{\gamma_1}}_{\beta_2}
    \Cdot {\VV^{\beta_2}}_{\beta_1} \,.
\label{F19}
\end{equation}
The equation \eref{F3} and the equations (9), (13), (14) from \cite{Ket} are
the particular cases of the last differential relation.

In the case of 1-dimensional functional dependence
$x^\alpha=x^\alpha(x^\beta)$ for two arbitrary coordinates $x^{\alpha}$
and $x^{\beta}$, \eref{F19} reduces to
\[
     \delta {\VV^{\alpha}}_{\beta} =
     \delta \psi \left[
          1\pm \left({\VV^{\alpha}}_{\beta}\right)^2
     \right] \,,
\]
where $\psi$ is turn angle in the plane $(\D x^{\beta},\D x^{\alpha})$,
and the sign of $(\VV^\alpha_\beta)^2$ is determined the signs of
metric tensor components $\g_{\alpha\alpha}$ and $\g_{\beta\beta}$. Its
solution is
\[
     {\VV^{\alpha}}_{\beta} =
     \cases{
          \tan \psi, &for $\g_{\alpha\alpha}= \g_{\beta\beta}$;\\
          \tanh\psi, &for $\g_{\alpha\alpha}=-\g_{\beta\beta}$.\\
     }
\]

\section{Conclusions}

The expansion of the space-time to the generalized space-time is the
consequence of uniform approach to the differentiation of kinematic
variables. In order to increase the space dimension, the following
recurrence rule is used:  in the next step of generalization, additional
coordinates of space-time are angles of turns in planes involving the
coordinates which were introduced in the previous step.  The
Minkowskian space-time is used for starting the recurrence
generalization procedure.

In context of the specified generalization, the new kinematic constants
are introduced for application.  For example, the only fundamental
constant such as the velocity of light $c$ is required in the first
step, but the two additional constants such as the fundamental radius
$R$ and time $T$ (or the fundamental angular velocity $\Omega$ and
acceleration $A$) are required in the second step.

We propose that the motion of light (as particle or wave)
is subject to the condition for the interval square in the generalized
space-time
\[
      (\D \sigma )^2 = 0\,.
\]
The kinematic properties of light becomes more rich through the expansion of
space-time.  In the Einsteinian special relativity, the light is in linear
uniform velocity motion with fundamental velocity $c$. For the second step of
generalization, the light motion includes accelerated motion and proper
rotation. This sequence can be continued. It is clear that the kinematics of
higher order can describe not only motions of new kinds but any motions
represented by the kinematics of the previous order as well.

The space generalization considered is naturally linked to a wide
spectrum of theoretical conceptions like Kaluza-Klein theory where
additional dimensions are invoked to describe different phenomena in
the uniform context.

\ack
The author is grateful to Dr.~A.~A.~Averyanov for helpful comments and
discussions.

\end{document}